\documentclass[twocolumn]{aastex701}

\usepackage{amsmath}

\begin{document}
\title{Periodicities of small regular jet deviations in SS433 and their origin}

\author[orcid=0000-0002-6755-2120,sname='Dodin']{A.V. Dodin}
\affiliation{Sternberg Astronomical Institute, Lomonosov Moscow State University, Universitetskij prospekt 13, 119234 Moscow, Russia}
\email[show]{samsebedodin@gmail.com}

\author[orcid=0000-0002-1705-617X, sname='Postnov']{K.A. Postnov}
\affiliation{Sternberg Astronomical Institute, Lomonosov Moscow State University, Universitetskij prospekt 13, 119234 Moscow, Russia}
\email[noshow]{kpostnov@gmail.com}

\author[orcid=0000-0001-5595-2285, sname='Cherepashchuk']{A.M. Cherepashchuk}
\affiliation{Sternberg Astronomical Institute, Lomonosov Moscow State University, Universitetskij prospekt 13, 119234 Moscow, Russia}
\email[noshow]{cherepashchuk@gmail.com}

\begin{abstract}
Based on the results of spectroscopic observations of SS433 conducted with the 2.5-m telescope at the Caucasian Mountain Observatory between 2020 and 2026, we study the deviations of the moving hydrogen and helium emission lines from their positions as predicted by the kinematic model of the jets’ precessional motion. In these deviations, in addition to the previously known periodicities of $13\fd08$, $11\fd27$, $6\fd29$, and $5\fd83$, a new period of $3\fd08$ is identified. All observed periodicities are explained by a model in which the jet offsets occur in a fixed direction relative to the star-disk axis $\bf r$, and the magnitude of the offset is a function $a(\mu)$, where $\mu$ is the cosine of the angle between $\bf r$ and the direction of each jet. The odd terms of the power series expansion of $a(\mu)$ in $\mu$ yield periodicities of $6\fd29$, $5\fd83$, and $3\fd08$, with the jets shifting in opposite directions while remaining parallel. The even powers yield periodicities of $13\fd08$, $11\fd27$, and $4\fd14$ (the latter is not observed), with both jets shifting in the same direction, which was previously misinterpreted as a change in the jets’ total velocity $\beta$. With the exception of the $6\fd29$ periodicity, all other displacements occur at an angle of $\approx120^\circ$ to $\bf r$. Superimposed on the regular variability are random deviations (jitter) in the jet directions, independent for the two jets, with $\sigma\lesssim1.\!\!^\circ5$.
\end{abstract}



\section{Introduction}

\label{sec:intro}
The unique object SS433 is the first galactic microquasar to be discovered. It is a massive X-ray binary system at an advanced stage of evolution with a precessing supercritical accretion disk around a black hole and relativistic jets \citep{Cher81a, Margon84, 2004ASPRv..12....1F, Cher20, 2025PhyU...68.1042C}.

The presence of precession means that the system has two periods: an orbital period ($P_{\rm Orb}\simeq13\fd1$), which is reliably determined from observations of both primary and secondary eclipses, and a precession period ($P_{\rm Prec}\simeq$ $162\fd3$), which is determined both from photometric observations and from spectral observations based on the positions of the moving hydrogen and neutral helium lines. All other reliably identified frequencies in the system are combinations of the orbital frequency $\omega=2\pi P_{\rm Orb}^{-1}$ and the precession frequency $\Omega=2\pi P_{\rm Prec}^{-1}$: the strongest deviations from the precessional motion occur with a period of $6\fd29$ ($2\omega+\Omega$). These deviations are commonly referred to as nutation and are associated with the star’s tidal action on the disk \citep{1982ApJ...260..780K}. The standard kinematic model \citep{Margon84} for spectral line shifts accounts only for precession and nutation variations. This model assumes that both jets move at speeds equal in magnitude but strictly in opposite directions:
\begin{equation}
\begin{split}
  z^{\pm}+1         &=z_{\rm Prec}^{\pm}+z_{\rm Nut}^{\pm };                \\
  z_{\rm Prec}^{\pm}      &=\gamma [  1 \pm \beta \sin i \sin \theta \cos \psi(t) \pm \beta \cos i \cos \theta ]        ;\\
  z_{\rm Nut}^{\pm}    &= \pm Z_{\rm Nut}\cos\phi_{\rm Nut}(t)                                        ;\\
  \psi(t)           &= 2\pi(t-T_3)/P_{\rm Prec};    \\
  \phi_{\rm Nut}(t) &= 2\pi(t-T_{\rm Nut})/P_{\rm Nut}.
\end{split} \label{eq:km}
\end{equation}
Here, $z^{\pm}=\lambda^{\pm}/\lambda_0-1$ is the observed line shift; $z_{\rm Prec}^{\pm}$ is the component that accounts for the transverse Doppler effect and precession; $z_{\rm Nut}^{\pm}$ is a correction accounting for nutation.
$\beta = v/c$ is the velocity of the material in the jet.
$i$ -- the inclination of the precession axis relative to the line of sight.
$\theta$ -- the aperture angle of the precession cone.
$\gamma = (1-\beta^2)^{-0.5}$ -- the Lorentz factor.
$P_{\rm Prec}$ -- the precession period,
$T_3$ -- the initial epoch when the wavelength difference between the moving lines is maximal (without accounting for nutation).
$Z_{\rm Nut}$ -- amplitude of the variation $z_{\rm Nut}^{\pm}$,
$P_{\rm Nut}$ -- nutation period, $T_{\rm Nut}$ -- initial epoch for describing nutation.

However, the moving lines do not strictly follow this model. At long timescales of $\sim100^{\rm d}$, fluctuations in the precession phase of $\sim\pm10$ days are noticeable \citep{2022ARep...66..451C}. On shorter timescales, random deviations in the jet direction are observed, with the angle changing independently for each jet \citep{1982ApL....23...11K, 2005ApJ...622L.129B}. Another periodicity of $5\fd8$ ($2\omega+3\Omega$) with a small amplitude is observed \citep{1982ApJ...260..780K, Davydov08}, which is not explained by the tidal nutation model \citep{1982ApJ...260..780K}. Furthermore, variations in $\beta$ with an orbital period have been reliably established \citep{2004ApJ...616L.159B, 2005ApJ...622L.129B, Blundell07, Davydov08, Cher18, 2022ARep...66..451C}, as well as with a period of $11\fd27$ $(\omega+2\Omega)$ \citep{Davydov08}.

Most position measurements of the moving lines in previous studies were carried out for H$\alpha,$ which is the strongest moving line in the visible range. However, this very feature is also a drawback: H$\alpha$ has a noticeable emission decay time, such that even a day later it may leave a bright residual trace. This trace can interfere with the measurement of a more recent component or even lead to its misidentification. Furthermore, H$\alpha$ lines can overlap with other lines in the stationary spectrum, making them unsuitable for precise measurements. The primary objective of this work was to enhance the precision of redshift determinations for the moving emission features, with the aim of obtaining a more detailed and statistically sound description of the stochastic deviations in jet orientation. This was accomplished through a simultaneous analysis of multiple H and He lines over an extended and internally consistent dataset of spectral observations.

\section{Observations}
The observations were carried out with the TDS spectrograph \citep{Potanin20} at the Caucasian Mountain Observatory of the Sternberg Astronomical Institute (SAI), Lomonosov Moscow State University, during 2020 -- 2026. The spectrograph’s operating wavelength range is $3550-7450${\AA}, with a resolving power of $\lambda/{\rm FWHM}\sim1000$ in the blue part of the spectrum and up to 2400 in the red part.
A total of 461 spectra were obtained, each formed by averaging three consecutive exposures of 5 minutes. Atmospheric extinction was corrected using a standard star observed at a similar airmass. For spectra acquired since 2022, this procedure includes telluric correction, which significantly improves the accuracy of measurements of the moving lines, when they overlap with telluric absorption bands. All timestamps and wavelengths were reduced to the barycenter of the Solar System.

\section{Parameters of the kinematic model in 2020 -- 2026}

The quality of the TDS spectra allows us to measure the redshifts $z^{\pm}$ not only for the moving H$\alpha^{\pm}$
lines, but also for higher members of the Balmer series up to H$\delta$, as well as for the He\,I lines at 5876, 6678, and 7065 {\AA}.  The set of measurable lines varies from spectrum to spectrum, depending on their availability for reliable measurement. The redshifts of all moving lines are consistent with one another to within  $\sigma_{z^\pm}=1.3\times10^{-3}.$ Given this level of agreement, we combine and average the measured $z^\pm$ values. This approach enhances both the accuracy and robustness of the $z^\pm$ measurements: if some moving lines are blended with other spectral features or fall outside the spectrograph’s wavelength range, $z^\pm$ can still be determined from the remaining measurable lines.

Building on our previous works \citep{Davydov08, Cher18, 2022ARep...66..451C} on determining the parameters of the kinematic model (Eq.~\ref{eq:km}), we used the same methods to determine the parameters from the new data, which are compiled in Tables \ref{tabl:prec} and \ref{tabl:nut}. The deviations of the observed line positions from the kinematic model with parameters from Tables \ref{tabl:prec} and \ref{tabl:nut} amount to $\sigma_{z^\pm}\approx5\times10^{-3},$ which is an order of magnitude greater than the measurement precision. In many cases, the model deviations are so large that they exceed the boundaries of the observed line profile, whose median FWHM is $v/c\approx6\times10^{-3}.$

\begin{table*}
\caption{Precession model parameters in different times.}
\label{tabl:prec}
\begin{tabular}{c | c | c | c | c | c | c  }
\hline
N & $\rm JD_{start}$ -- $\rm JD_{end},$ & $\beta$ & $\theta,$ & $i,$ & $P_{\rm prec},$     & $T_3,$              \\ 
  &           days                     &         &   $^o$    & $^o$ & days               & MJD               \\ 
\hline                                                                                                              %
1  & 58895 --  59044 &  $0.2655\pm0.0025$ &  $18.88\pm0.47$ & $78.34\pm0.24$ & $160.903\pm1.257$ &  $59096.7\pm1.2$ \\
2  & 59081 --  59202 &  $0.2604\pm0.0020$ &  $20.04\pm0.26$ & $78.84\pm0.21$ & $166.266\pm1.818$ &  $59260.3\pm1.2$ \\
3  & 59257 --  59560 &  $0.2619\pm0.0025$ &  $19.67\pm0.31$ & $79.50\pm0.22$ & $160.800\pm0.671$ &  $59579.0\pm0.7$ \\
4  & 59630 --  59795 &  $0.2625\pm0.0025$ &  $19.57\pm0.34$ & $78.37\pm0.23$ & $158.372\pm1.063$ &  $59895.3\pm1.2$ \\
5  & 59840 --  60002 &  $0.2671\pm0.0019$ &  $18.27\pm0.26$ & $78.99\pm0.22$ & $158.657\pm1.189$ &  $60055.3\pm1.2$ \\
6  & 60002 --  60254 &  $0.2536\pm0.0016$ &  $20.71\pm0.21$ & $78.73\pm0.14$ & $165.761\pm0.531$ &  $60223.1\pm0.3$ \\
7  & 60254 --  60497 &  $0.2563\pm0.0036$ &  $19.87\pm0.51$ & $79.08\pm0.34$ & $164.932\pm1.680$ &  $60387.6\pm0.8$ \\
8  & 60496 --  60655 &  $0.2524\pm0.0029$ &  $20.29\pm0.41$ & $78.54\pm0.32$ & $166.066\pm1.464$ &  $60719.5\pm1.8$ \\
9  & 60718 --  60891 &  $0.2623\pm0.0026$ &  $18.94\pm0.38$ & $78.17\pm0.31$ & $157.420\pm1.638$ &  $60871.0\pm0.8$ \\
10 & 60891 --  61076 &  $0.2511\pm0.0024$ &  $22.06\pm0.49$ & $78.72\pm0.39$ & $166.632\pm1.690$ &  $61037.9\pm0.8$ \\
11 & 61076 --  61247 &  $0.2605\pm0.0026$ &  $19.00\pm0.37$ & $78.65\pm0.26$ & $162.758\pm1.101$ &  $61199.1\pm0.4$ \\
\hline
All$^*$& 43688 --  61247 &  $0.2598\pm0.0007$ &  $19.60\pm0.09$ & $78.98\pm0.06$ & $162.309\pm0.002$ &  $60225.88\pm0.16$ \\
\hline
\multicolumn{7}{r}{$^*$Parameters in the last row are based on the full 1978 -- 2026 dataset \citep{Cher18, 2022ARep...66..451C}.}

\end{tabular}
\end{table*}

\begin{table}
\caption{Nutation model parameters.} \label{tabl:nut}
\begin{tabular}{c|c|c|c}
\hline
N  &  $Z_{\rm Nut},$  & $P_{\rm Nut},$   & $T_{\rm Nut},$ \\ 
   &  $10^{-3}\times$  & days           &  days          \\ 
\hline
1  & $ 8.1\pm1.1$  & $6.335\pm0.028$ &  $58936.54\pm0.15$ \\
2  & $ 9.2\pm0.8$  & $6.314\pm0.019$ &  $59125.02\pm0.12$ \\
3  & $10.4\pm1.0$  & $6.282\pm0.008$ &  $59445.48\pm0.10$ \\
4  & $ 7.8\pm1.1$  & $6.304\pm0.018$ &  $59747.38\pm0.16$ \\
5  & $ 9.7\pm0.9$  & $6.346\pm0.011$ &  $59936.04\pm0.09$ \\
6  & $10.2\pm0.7$  & $6.276\pm0.006$ &  $60149.72\pm0.06$ \\
7  & $11.4\pm1.6$  & $6.242\pm0.017$ &  $60495.04\pm0.24$ \\
8  & $ 8.5\pm1.3$  & $6.312\pm0.017$ &  $60495.43\pm0.20$ \\
9  & $11.0\pm1.3$  & $6.255\pm0.019$ &  $60809.61\pm0.13$ \\
10 & $ 6.4\pm1.1$  & $6.315\pm0.030$ &  $60809.17\pm0.77$ \\
11 & $ 8.8\pm1.0$  & $6.232\pm0.020$ &  $60813.40\pm1.29$ \\
\hline
\end{tabular}
\end{table}

\section{Deviations from the model}
\label{sect:dev}

\begin{figure}
\centering
\includegraphics[width=\hsize]{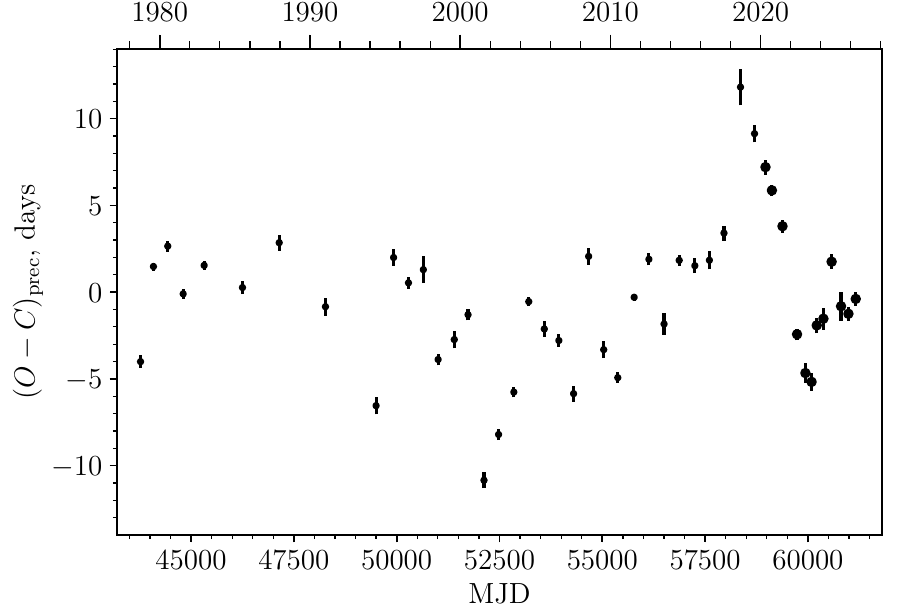}
      \caption{The precession phase deviation from the mean ephemeris. The analyzed period is highlighted with larger markers.}
         \label{fig:ochz}
\end{figure}

We deliberately divided the entire observed period of 2020 -- 2026 into the shortest possible intervals that still allow us to determine the model parameters with acceptable accuracy. As can be seen from Table~\ref{tabl:prec}, the precession period exhibits fluctuations. This effect is interpreted as phase noise in the precessional variability, likely caused by fluctuations in the viscous timescale in the disk (see \citealt{2022ARep...66..451C}). Using the same methods, we extended the time series of phase shifts from \citet{2022ARep...66..451C} with the new data, see Fig.\,\ref{fig:ochz}. Since the precessional component of the variations has the largest amplitude, all other small‑scale deviations will be studied after subtracting the precessional contribution with parameters from Table~\ref{tabl:prec}.

Our kinematic model incorporates nutational variability, which in principle could be fitted on a segment‑by‑segment basis using the parameters listed in Table~\ref{tabl:nut}. However, to preserve long‑term components that may be physically significant, we deliberately avoid fitting the nutation parameters over short time intervals: such an approach risks suppressing slow variations whose characteristic timescales exceed the length of the individual segments. Instead, the nutational component is determined globally across the entire 2020 -- 2026 dataset. This choice ensures that any long‑term trends or low‑frequency signals are retained in the residuals, which are the primary focus of our analysis. We note, however, that the main results remain consistent even if the nutational contribution is subtracted in advance using the global parameters from Table~\ref{tabl:nut}, confirming the robustness of our conclusions.

\subsection{Periodicities in the jet velocity $\beta$}
\label{sect:beta}
The kinematic model Eq.~\eqref{eq:km} does not account for the variability of the jet velocity $\beta=v/c.$
Nevertheless, such variability with an orbital period was detected by \cite{2005ApJ...622L.129B} and has been repeatedly confirmed thereafter by \cite{Blundell07, Davydov08, Cher18, 2022ARep...66..451C}. Since the orbital period is slowly increasing \citep{2023NewA..10302060C}, hereafter we adopt its mean value over the 2020 -- 2025  interval, $P_{\rm Orb}=13\fd08341,$ and use the epoch of the primary eclipse at the midpoint of this interval, ${\rm MJD} = 58932.636,$ to compute the orbital phase.

\begin{figure}
\centering
\includegraphics[width=0.75\hsize]{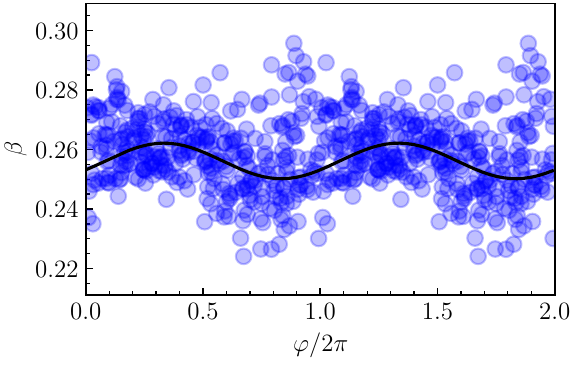}
      \caption{Variation in the jet velocity with orbital phase.}
         \label{fig:beta1308}
\end{figure}

Assuming that the velocities in both jets are equal in magnitude and strictly opposite in direction, the velocity $\beta$ can be computed for each individual observation from the transverse Doppler effect:
\begin{equation}\label{betaEq}
\beta=\sqrt{1- \left(1+\frac{z_{-}+z_{+}}{2} \right)^{-2}}.
\end{equation}
Assuming such $\beta$ measurements, we detect a period of $13\fd093 \pm 0\fd005,$
which differs by $0\fd009$  from the orbital period. This discrepancy may be attributed to strong contamination of the curve by spurious effects, for example, due to the failure of the strict antiparallel alignment of the jets. This appears to be the primary cause of irregular deviations of the line positions from the model, see Section \ref{sect:irreg}. For definiteness, we assume that the period in the $\beta$ variation is exactly equal to the orbital period; the remaining parameters were determined via the least‑squares method and are listed in Table\,\ref{tabl:beta}. The dependence of $\beta$ on the orbital phase is shown in Fig.\,\ref{fig:beta1308}; the velocity maximum occurs at $\varphi=0.334\pm0.015.$ We note that all subsequent conclusions remain valid if we adopt the period of $13\fd093$ instead.
\begin{table}
\caption{Periodical components in the jet velocity. The mean $\beta=0.2562\pm0.0004.$} \label{tabl:beta}
\begin{tabular}{c|c|c|c}
\hline
$P,$ days &$\Delta z,$ $\times10^{-3}$ & $\Delta\beta,$ $\times10^{-3}$      &$\rm MJD_0$         \\
\hline
13.083 &1.75$\pm$0.18 &6.0$\pm$0.6 & 58937.0$\pm$ 0.2 \\
11.267 &1.32$\pm$0.15 &4.5$\pm$0.5 & 58895.1$\pm$ 0.2                  \\
\hline
\end{tabular}
\end{table}

\begin{figure}
\centering
\includegraphics[width=0.75\hsize]{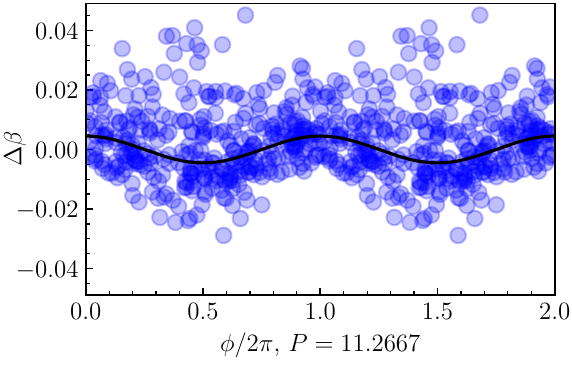}
      \caption{A $11.267$-day periodicity in the jet velocity residuals after subtracting the orbital variability from Fig.\,\ref{fig:beta1308}.
}
         \label{fig:beta1126}
\end{figure}

After removing the orbital component from $\beta$, the strongest harmonic in the variability becomes $P=11\fd251\pm0\fd004$, which is close to the expected harmonic $11\fd267$ corresponding to the combination frequency
$(\omega+2\Omega)$. As before, we adopt the theoretically expected period of $11\fd267$ for further analysis, although the subsequent results are practically insensitive to this choice. The remaining parameters of this periodicity are listed in Table\,\ref{tabl:beta}, and the phase curve is shown in Fig.\,\ref{fig:beta1126}.

The periodic components $\beta$ found in this way were substituted into the kinematic model (Eq.~\ref{eq:km}) using the parameters from Tables \ref{tabl:prec}, \ref{tabl:nut}, and the differences between the model with a variable $\beta$ and the model with a constant $\beta$ were calculated. The resulting differences were used to correct the observed $z^{\pm}$ for the periodic variation of the velocity $\beta.$


\subsection{Periodic Changes in Jet Direction}
\begin{figure*}
\centering
\includegraphics[width=\hsize]{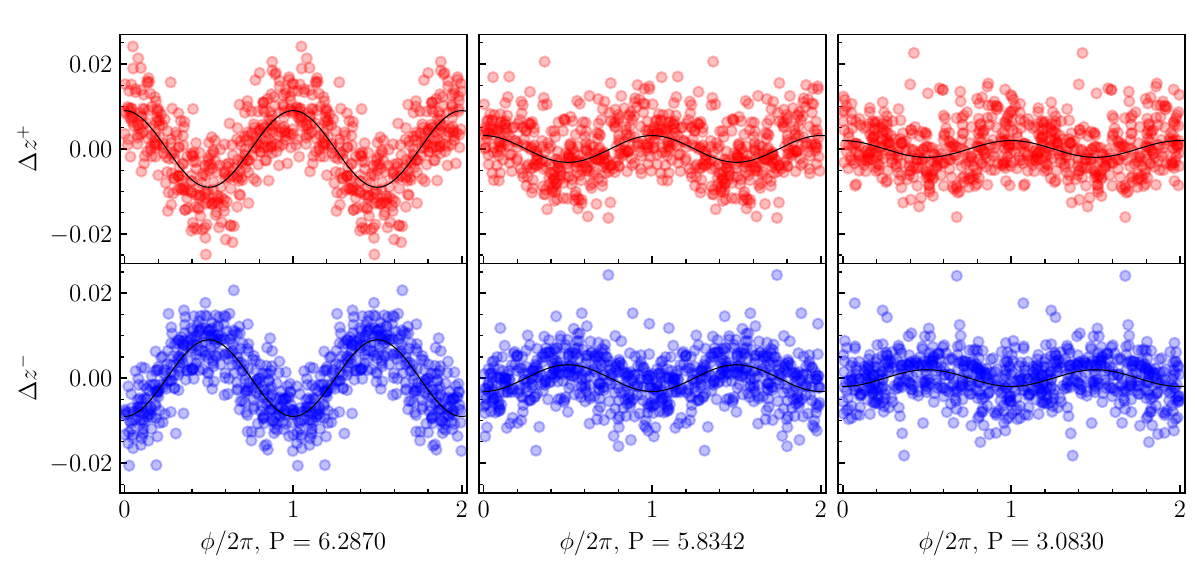}
      \caption{Periodicities in the residual deviations $z^{\pm}$. The precession component and the periodic components $\beta$ have been subtracted beforehand.}
         \label{fig:pjitter}
\end{figure*}

After removing corrections for periodic variations in $\beta$ from $z^{\pm}$, all periodic components that are simultaneously present in both jets were sequentially extracted from the data, see Fig.\,\ref{fig:pjitter}. In addition to the well-known nutation, the variations contain a period of $5\fd8342$ days, which was also found in papers \citet{1982ApJ...260..780K, Davydov08} and corresponds to the harmonic $(2\omega+3\Omega)$, and a period of $3\fd0830$ days, not previously mentioned in the literature and corresponding to the harmonic $4\omega+3\Omega.$ The parameters of these periodicities are summarized in Table \ref{tabl:tab3}.

\begin{table}
\caption{Periodic components in the deviations of moving lines from the precession model.} \label{tabl:tab3}
\begin{tabular}{c|c|c}
\hline
 $P,$ days         & $\Delta z$, $\times10^{-3}$  &  $T_0, $ MJD      \\
\hline
6.2870$\pm$0.0003 & 9.0$\pm$0.3     &  60005.14$\pm$0.03\\
5.8342$\pm$0.0008 & 3.2$\pm$0.3     &  60003.98$\pm$0.09\\
3.0830$\pm$0.0003 & 2.0$\pm$0.3     &  60006.72$\pm$0.07\\
\hline
\end{tabular}
\end{table}

There are other peaks present in the frequency spectra; however, they do not coincide for the red $z^+$ and blue $z^-$ jets, nor do they correspond to combinations of known frequencies, so we consider them to be noise.

\subsection{Aperiodic components}
\label{sect:irreg}
\begin{figure}
\centering
\includegraphics[width=\hsize]{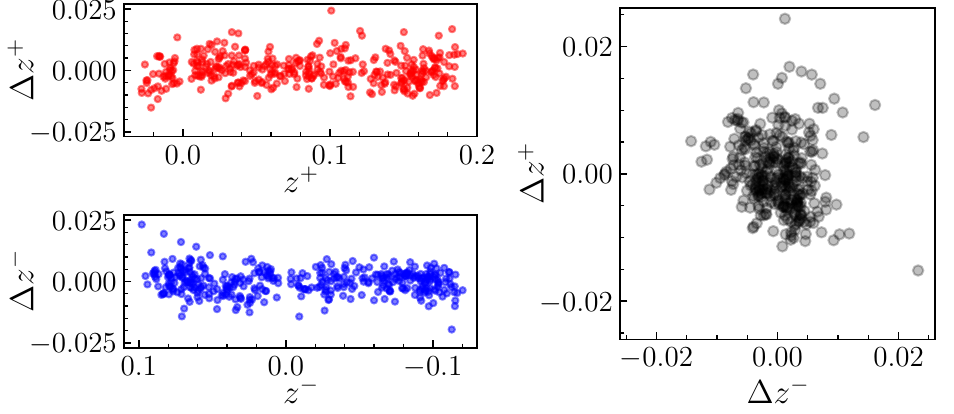}
      \caption{Residual deviations $\Delta z^{\pm}$ after eliminating all periodic components, as a function of the observed $z^{\pm}$ and relative to each other.}
         \label{fig:zdz}
\end{figure}

The regular deviations from the precession-nutation model given in Eq.~\eqref{eq:km}, discussed above, account for only a small portion of the total dispersion: after subtracting all periodic components, the dispersion decreases from $\sigma_{z^-}=5.3\times10^{-3}$ and $\sigma_{z^+}=6.3\times10^{-3}$ to $\sigma_{z^-}=5.0\times10^{-3}$ and $\sigma_{z^+}=5.6\times10^{-3}.$ The residuals $\Delta z^{\pm}$ are shown in Fig.\,\ref{fig:zdz} as functions of $z^{\pm}$ and of each other; in the latter case, a slight elongation of the scatter plot is noticeable, which corresponds to an anticorrelation between $\Delta z^{\pm}$ with a correlation coefficient of $r=-0.19\pm0.05.$

\begin{figure}
\centering
\includegraphics[width=\hsize]{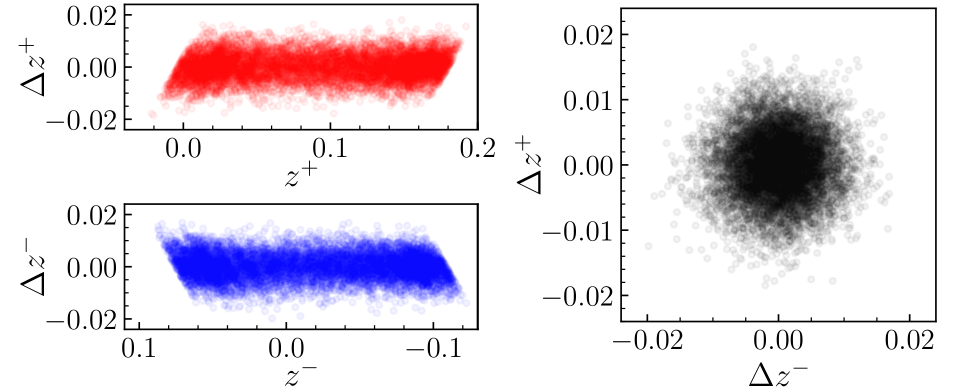}
      \caption{Expected correlations $\Delta z^{\pm} - z^{\pm}$ and $\Delta z^{+} - \Delta z^{-}$ for random deviations in the directions of each jet with $\sigma = 1.5$ degrees (diameter of the cone).}
         \label{fig:2angle}
\end{figure}

\begin{figure}
\centering
\includegraphics[width=\hsize]{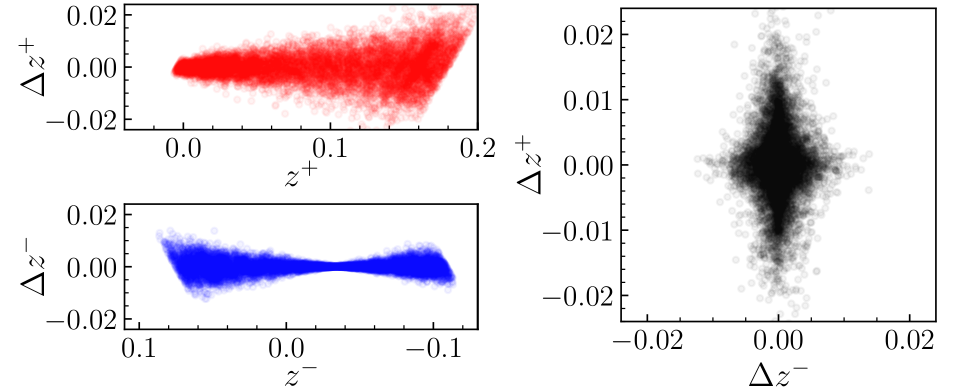}
      \caption{
      Expected correlations $\Delta z^{\pm} - z^{\pm}$ and $\Delta z^{+} - \Delta z^{-}$ for random velocity variations $\beta$ of each jet with $\sigma = 0.01.$}
         \label{fig:2beta}
\end{figure}

The diagrams shown narrow down the range of causes that lead to deviations in the lines \citep{1982ApL....23...11K, 2005ApJ...622L.129B}. The observed pattern of variations corresponds to random and independent variations in the directions of the jet and counterjet, see Fig.\,\ref{fig:2angle}. If the deviations occurred only in the total angle, that is, if both jets and the source lay strictly on a single straight line, then $\Delta z^{\pm}$ would be strictly anticorrelated, and the cloud of points on the diagram $\Delta z^- - \Delta z^+$ would degenerate into a straight line. With random variations in $\beta$ independent in each jet, there should be a strong dependence of the deviations $\Delta z^{\pm}$ on $z^{\pm}$, see Fig.\,\ref{fig:2beta}. Such a dependence is not observed.

\begin{figure}
\centering
\includegraphics[width=\hsize]{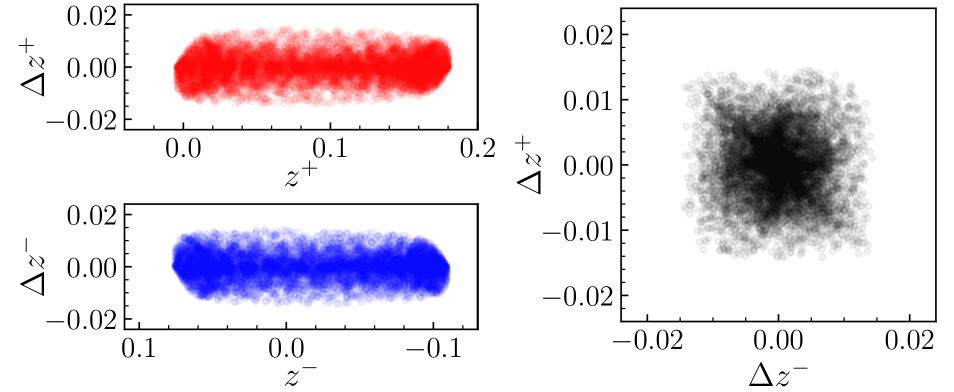}
      \caption{Expected relationships between $\Delta z^{\pm} - z^{\pm}$ and $\Delta z^{+} - \Delta z^{-}$ for random changes in delays uniformly distributed over the interval $\pm 1$ day.}
         \label{fig:2tau}
\end{figure}

Another factor that may lead to deviations from the model is the unknown age of the moving lines. As the emitting gas moves along a ballistic trajectory, it fixes the position of the jet at the moment of its launch, but an unknown time interval separates the moment of launch from the moment of observation. Since the kinematic model is constructed based on the observed lines, it incorporates an average delay, whereas the delays of individual lines will be both shorter and longer than this average delay. If the delays in the jet and counterjet lines are identical, there must be a strict anticorrelation $\Delta z^{\pm}$. In the case where the delays in the jets are independent of one another, both diagrams will resemble the observed ones, see Fig.\,\ref{fig:2tau}.

Yet another effect that could distort the radial velocity curve of moving lines is the light delay between the receding and approaching jets. However, this effect does not need to be taken into account. The fact is that we observe the jet not as a continuous stream, but as separate clumps that travel along a ballistic trajectory and carry information about their radial velocity at the moment of launch. Therefore, the only important factors are whether we are observing clumps launched simultaneously and how long ago this occurred.

In our view, the most plausible cause of the deviations is random fluctuations in the directions of each jet. The scatter $\sigma_{z^-}=5\times10^{-3}$ corresponds to a spread cone diameter of $\sigma=1.\!\!{^\circ}5.$ The slight elongation of the point cloud in the $\Delta z^+ - \Delta z^-$ plot (Fig.\,\ref{fig:zdz}) is due to the uncertainty in the age of the observed moving lines, as well as a possible weak correlation between the angular variations in the jets. Taking into account the contributions of various mechanisms to the total scatter, we can state that the angular fluctuations satisfy $\sigma\lesssim1.\!\!{^\circ}5.$

\section{Discussion}
\begin{figure}
\centering
\includegraphics[width=\hsize]{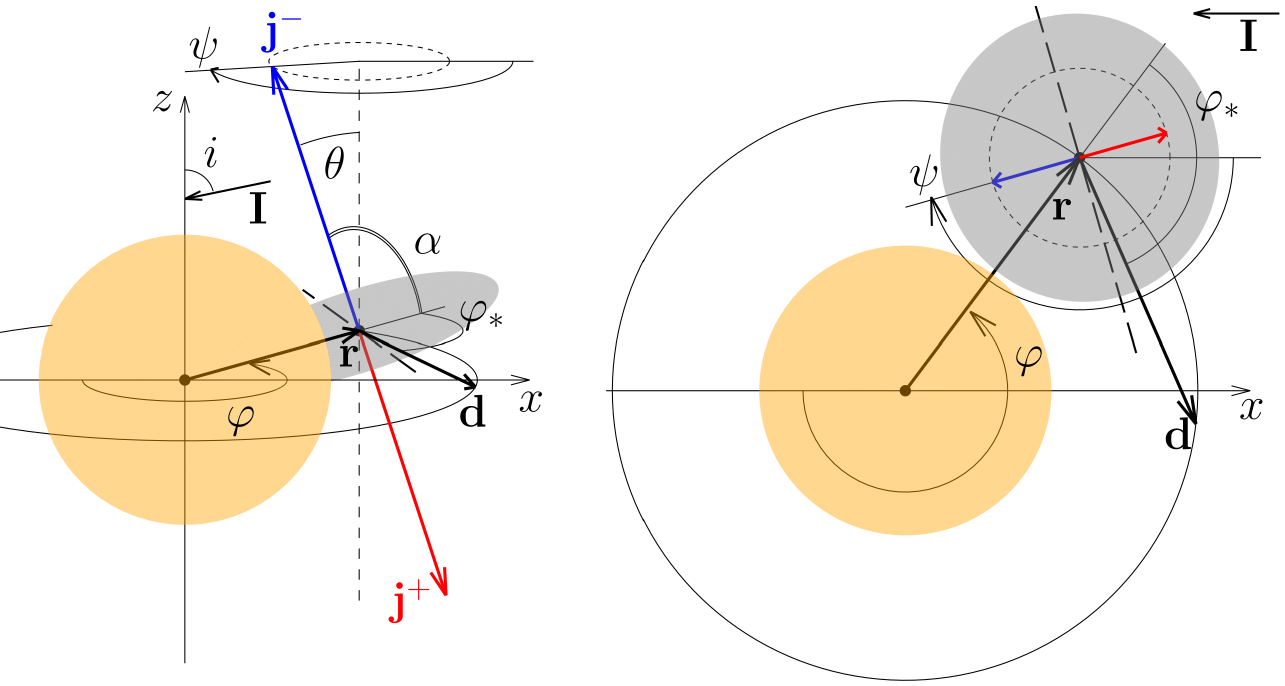}
      \caption{Schematic representation of SS433. On the left, a view of the system at $\varphi = 90^\circ$ and at an inclination $i=79^\circ.$ The disk and jet positions correspond to $\varphi/2\pi=0.65$ and $\psi/2\pi=0.46.$ On the right is the top view of the system. The bold dashed line is the intersection line between the disk plane and the orbital plane. The angle between the jet $j^{-}$ and the radius vector $\bf r$ connecting the centers of the components is $\alpha;$
      $\mu_{-}=\cos\alpha.$}
         \label{fig:vec}
\end{figure}

To interpret the detected periodicities, we need to formalize the basic elements of the system, which is conveniently done in vector form (see Fig.\,\ref{fig:vec}). The position vector from the star to the disk, lying in the orbital $xy$ plane, is given by:
\begin{equation}
\begin{split}
{\bf r} &= -\{\cos\varphi, \sin\varphi, 0 \}, \label{eq:rorb}
\end{split}
\end{equation}
The sign `$-$' means that $\varphi$ is traditionally counted from the moment of eclipse of the disk by the star $T_{\rm Orb}$.
To calculate $\varphi,$ we used the average period for our observation interval (see. \citealt{2023NewA..10302060C} and \citealt{2026ApJ...999..104D}):
\begin{equation}
\begin{split}
&\varphi=\omega (t-T_{\rm Orb}), \quad \omega=\frac{2\pi}{P_{\rm Orb}}, \\
&T_{\rm Orb} = 58932.636\,{\rm MJD}, \quad  P_{\rm Orb} = 13\fd08341.
\end{split}
\end{equation}
The precessional phase $\psi$ is calculated from the data in Table~\ref{tabl:prec}:
\begin{equation}
\begin{split}
&\psi=\Omega (t-T_{3}), \quad \Omega=\frac{2\pi}{P_{\rm Prec}},\\
&T_3 = 60225.88\, {\rm MJD}, \quad P_{\rm Prec} = 162\fd309.
\end{split}
\end{equation}
The disk in SS433 precesses retrogradely with respect to the orbital motion; for convenience, we also measure $\psi$ in the opposite direction to keep $\Omega$ positive.

The jets are precessing around the orbital angular momentum under the angle $\theta$ to the $z$-axis:
\begin{equation}
\begin{split}
{\bf j^\pm} = \mp\{\cos\psi\sin\theta, -\sin\psi\sin\theta, \cos\theta \}.
\end{split}
\end{equation}
The sign `$\mp$' correspond to the definition that the phase  $\psi=0$ corresponds the maximum disk opening time $T_3$. The sign `$-$' before `$\sin\psi$' reflects the chosen direction of $\psi.$

The line of sight defines the plane  $xz$ and is inclined under the angle $i$ to the orbital angular momentum:
\begin{equation}
{\bf I} = -\{\sin i, 0, \cos i\}.
\end{equation}
The signs correspond to the definition that radial velocity of receding objects is positive. The radial velocity of the jet is determined by its projection on the line of sight:
\begin{equation}
\eta^{\pm}={\bf j^\pm\cdot I} = \pm(\cos\psi\sin\theta\sin i+\cos\theta\cos i),
\label{eq:jet}
\end{equation}
which with account of the relativistic factors yields the precessional part of the kinematic model (Eq.~\ref{eq:km}):
\begin{equation}
z^{\pm} = \gamma(1+\beta\eta^{\pm})-1.
\label{eq:zpm}
\end{equation}
To interpret the deviations of the jet from the precessional motion, we consider the torques acting on the disk due to the star and the gas flows in the system. The dominant flow is the one feeding the disk through the L$_1$ point. The mutual orientation of the gas flows and the precessing disk and star is determined by the co-addition of the orbital and precessional phases, $\varphi + \psi.$ Consequently, the torque on the disk should be a periodic function of $\varphi + \psi.$ The simplest form of such a function is a cosine, generally with some phase offset $\psi_*$: $\cos(\varphi+\psi-\psi_*).$
It is also natural to assume that the strength of this torque is proportional to the half‑angle of the precession cone, i.e., $\sin(\theta),$ as in the absence of precession there would be no deviations, and the gas flow pattern would be static in the frame corotating with the orbital motion. Note that, up to a phase, the product $\mu=\sin(\theta)\cos(\varphi+\psi-\psi_*)$ is equal to the cosine of the angle between the position vector ${\bf r}$ and the jet direction vectors ${\bf j^\pm}$:
\begin{equation}
\mu_{\pm}={\bf r\cdot j^\pm} = \pm\sin\theta\cos(\varphi+\psi).
\end{equation}
In general, the magnitude of the jet’s deviation from the precessional motion may be a nonlinear function $a(\mu),$ which remains periodic in $\varphi + \psi.$
Assuming that deviations from linearity are small, we can expand this function as a power series in $\mu:$
 \begin{equation}
a(\mu) = a_0+a_1\mu+a_2\mu^2+a_3\mu^3+\ldots.
\end{equation}
In practice, truncating the expansion at the quadratic or cubic term often suffices to capture the observed deviations while keeping the model analytically tractable.

The quantity $a(\mu)$ determines only the amplitude of the deviation; the direction of this deviation can conveniently be characterized by its projection in the orbital plane, offset from the radius vector $\bf r$ by a constant angle $\varphi_*$:
\begin{equation}
{\bf d} = -\{ \cos(\varphi-\varphi_*), \sin(\varphi-\varphi_*), 0 \}.
\end{equation}
New direction of the jet is
\begin{equation}
{\bf j^\prime} = \frac{{\bf j}+a(\mu){\bf d}}{|{\bf j}+a(\mu){\bf d}|},\label{eq:jprime}
\end{equation}
hereinafter, we omit the designation `$\pm$' for the jet and counterjet. Normalization means that we assume only the direction of the jet changes, while its speed remains unchanged; therefore, the new vector remains an unit vector. However, omitting this normalization does not affect the result, since we assume that the deviation is small and expand the denominator to the term linear in $a$. In this approximation, the change in radial velocity will be
\begin{equation}
{\Delta z} = \gamma \beta a(\mu)[\bf (d{\cdot}I)-(j{\cdot}I)(d{\cdot}j)]. \label{eq:dz}
\end{equation}
The last term can also be neglected, since it consists of periodic terms with amplitudes
$\sin^2\theta\approx0.1$ and $\sin\theta\cos i\approx0.06$ relative to the first term. Thus, the dominant term is
\begin{equation}
\Delta z = \gamma\beta a(\mu){\bf d\cdot I} = \gamma\beta\sum a_k\mu^k\cos(\varphi-\varphi_k)\sin i.
\end{equation}

Consider each term in this expansion separately.
For $k=0,$ the jet deviation does not depend on $\varphi+\psi$, and the velocity variation can occur only with the orbital phase. Since the deviation is independent of the sign of $\mu$, it is the same for the jet and the counterjet and occurs in the same direction; therefore, they cease to be antiparallel. This leads to an apparent variation of $\beta$ with the orbital period, since its calculation in the kinematic model (see Eq.\,\ref{betaEq}) is based on the assumption that the jet and counterjet are strictly antiparallel. Indeed, we observe a variation of $\beta$ with the orbital period (see Fig.\,\ref{fig:beta1308}). The orbital motion of the disk around the star should also contribute to the velocity variation with the orbital period at the level of $\Delta z\sim 2\times10^{-4}$; however, this is comparable to the accuracy with which the amplitude of the orbital variability of $z$ is determined.

For $k=1$, the deviation of the radial velocity from the precessional motion takes the form:
\begin{equation}
\begin{split}
\Delta z_1 = \gamma\beta a_1\mu{\bf d\cdot I} = & \gamma\beta\frac{a_1}{2}\sin i \sin \theta [
\cos (\psi +\varphi_1-\psi_1)+ \\
&\cos(2\varphi+\psi-\varphi_1-\psi_1)].
\label{eq:mu1}
\end{split}
\end{equation}
This law contains two characteristic frequencies: the precession frequency $\Omega$ and the nutation frequency $2\omega+\Omega.$ The precessional component is absorbed into the overall precession. Note that the sign of $\mu$ is opposite for the jet and counterjet, so their deviations occur in opposite directions, as observed in the nutational component (see the left panels of Fig.\,\ref{fig:pjitter}).

For $k=2$ the deviation of the radial velocity from the precessional motion reads
\begin{equation}
\begin{split}
\Delta z_2 = \gamma\beta a_2\mu^2{\bf d\cdot I} = &\gamma \beta \frac{a_2}{4}\sin i \sin^2 \theta[
       2\cos(\varphi -\varphi_2)+ \\
&       \cos(\varphi+2\psi+\varphi_2-2\psi_2)+\\
&       \cos(3\varphi+2\psi-\varphi_2-2\psi_2)].
\end{split}
\label{eq:mu2}
\end{equation}
This gives the frequencies $\omega$, $\omega+2\Omega$, and $3\omega+2\Omega$, corresponding to periods of $13\fd08$, $11\fd27$, and $4\fd14$, respectively.
As in the case of $k = 0$, the sign of $\mu^2$ does not depend on of whether we consider the jet or the counterjet; therefore, their deflections will be in the same direction, which may lead to an incorrect interpretation of the variability of the total speed $\beta.$

Indeed, we observe variability in $\beta$ with periods of $13\fd08$ and $11\fd27$ (see Section~\ref{sect:beta} above and Fig.\,\ref{fig:beta1308} and \ref{fig:beta1126}). We were unable to identify the $4\fd14$ period. In this connection, we note that detecting short periodicities is more difficult due to the comparable fading time of the moving lines, which is about 1 day.

For $k=3$ deviations of the radial velocity from the precessional motion are
\begin{equation}
\begin{split}
\Delta z_3 = \gamma \beta a_3\mu^3{\bf d\cdot I} = &\gamma \beta \frac{a_3}{8}\sin i \sin^3 \theta \times \\
\times[  3&\cos(\psi +\varphi_3-\psi_3)+     \\
         3&\cos(2\varphi+\psi-\varphi_3-\psi_3)+\\
          &\cos(2\varphi+3\psi+\varphi_3-3\psi_3)+    \\
          &\cos(4\varphi+3\psi-\varphi_3-3\psi_3)].
\end{split}
\label{eq:mu3}
\end{equation}
This yields, once again, the precession and nutation frequencies, as well as the frequencies $2\omega+3\Omega$ and $4\omega+3\Omega,$ corresponding to the observed periodicities of $5\fd83$ and $3\fd08$, respectively.

The unknown parameters of Eqs.~\eqref{eq:mu1}-\eqref{eq:mu3} can be determined from observations (see Tables \ref{tabl:beta} and \ref{tabl:tab3}). Substituting $\varphi(T_0)$ and $\psi(T_0)$ for the initial epochs $T_{0}$ of each periodicity into the corresponding terms of Eqs. \eqref{eq:mu1}-\eqref{eq:mu3} yields linear equations for $\varphi _{k}$ and $\psi _{k},$ while the quantities $a_{k}$ are determined from the observed amplitudes $\Delta z.$

From Eq.~\eqref{eq:mu2} for the periods of $13\fd08,$ $11\fd27$ we obtain
\begin{equation}
\begin{split}
&\varphi_2/2\pi=0.334\pm0.015,   \\
&\psi_2/2\pi=0.033\pm0.011+m/2,
\end{split}
\label{eq:phi2}
\end{equation}
where $m$ is an integer.
Note that the phase delay $\varphi_{2}$ is directly observed in the periodicity $13\fd08$ in Fig.\,\ref{fig:beta1308}.
Matching the observed amplitudes with Eq.~\eqref{eq:mu2} yields
$a_2=0.139\pm0.014$ for the period $13\fd08,$ and $a_2=0.21\pm0.02$ for the period $11\fd27.$

Eq.~\eqref{eq:mu3} for the periods of $5\fd83$ and $3\fd08$ yields
\begin{equation}
\begin{split}
&\varphi_3/2\pi=0.330\pm0.013+n/2, \\
&\psi_3/2\pi=0.000\pm0.004+m/6,
\end{split}
\label{eq:phi3}
\end{equation}
where $n, m$ are integers. The coefficient $a_3=3.3\pm0.3$ and $a_3=2.1\pm0.3,$ respectively from the amplitudes of $5\fd83$ and $3\fd08$ periodicities.

If we set $n,m = 0,$ the obtained results acquire a clear physical meaning.
In this case all four periodicities share a common phase offset
$\varphi_{2,3}\approx0.33$ and $\psi_{2,3}\approx0,$ meaning that the jet deviations are proportional to the cosine of the angle between the jet and the position vector, $\mu_{\pm}$.

Both terms $k=1$ and $k=3$ contribute to the strongest $6\fd29$ periodicity. However, it turned out that Eq.~\eqref{eq:mu1} for $k = 1$ yields $(\varphi_1+\psi_1)/2\pi=0.589\pm0.005,$ which is inconsistent with the values of $\varphi_3$ and $\psi_3.$ To extract the contribution of $\mu^1$ from the periodicity $6\fd29$, the addition formula for cosines with different amplitudes and a phase shift should be used. The resulting equations for the amplitude and phase at $a_3=2.1\div3.3$ and $\varphi_3/2\pi=0.33$ yield $a_1=0.28\div0.33$ and $(\varphi_1+\psi_1)/2\pi=0.68\div0.71.$ The phase difference between linear and non-linear terms may be attributed to the fact that the linear term arises from a mechanism of a different nature than the nonlinear contributions. For example, the tidal influence of the star may dominate the linear term, while the nonlinear components are primarily driven by the effects of gas flows in the system.

The same reasoning can be applied to $k=0$. The agreement of the phases $\varphi_2\approx\varphi_3\approx0.33$ and $\psi_2\approx\psi_3\approx0$ implies that the contribution of the $k=0$ term to the $13\fd08$ periodicity is either very small or has $\varphi_0=0.33$ as well, since otherwise it would distort the values of $\varphi_2$ and $\psi_2$, and these would not agree with $\varphi_3$ and $\psi_3$. The smallness of $a_0\approx0$ is not surprising, since this term corresponds to a constant displacement of the jets in one direction relative to $\mathbf{r}$, independent of the precessional orientation of the disk.

The same problem can also be solved numerically in a more general form, without the expansion of Eq.~\eqref{eq:jprime} in the small parameter $a(\mu)$. Using the mean parameters of the kinematic model and the data from Tables~\ref{tabl:beta} and \ref{tabl:tab3}, we generated artificial observations and fitted them with a model given by Eqs.~\eqref{eq:rorb}--\eqref{eq:jprime}. We searched for a $\psi_k$ common to all $k$ and obtained $\psi_k/2\pi=0.02$, which is very close to zero and justifies setting $\psi_k=0$. We also searched for a common phase shift $\varphi_{2,3}$ for all periodicities except the component from $\mu^1$. This fit yielded the following values: $a_1=0.29$, $a_2=0.13$, $a_3=2.0$, $\varphi_1=0.70$, $\varphi_{2,3}=0.32$. Using this model, we also estimated the contribution of the precessional term from the odd powers of $\mu$ to the parameters of the kinematic model. The only parameter exhibiting a significant change is the angle $\theta:$ the standard kinematic model underestimates this angle by approximately $2^\circ$ relative to the true value. This shift is also accompanied by a delay of about two days in the epoch $T_3$ relative to the model prediction.

Adding to this model the noise in the direction of each jet, as found in Section~\ref{sect:irreg}, does not change the amplitudes of the periodicities. To mimic the uncertainty in the ages of the moving lines, we introduced random deviations in the time stamps, the same for both jets (see the conclusions of Section~\ref{sect:irreg}). Such phase noise reduces the amplitudes of the periodicities, the shorter the period, the stronger the reduction. Possibly for this reason, we see the $11\fd27$ periodicity in $\beta$ but do not detect the $4\fd14$ periodicity, although their amplitudes should be equal. The same may explain the decrease in the amplitude of the $3\fd08$ periodicity relative to that of $5\fd83$. Thus, for Gaussian noise with $\sigma=0\fd6$, the observed ratio of the amplitudes of these periodicities can be reproduced, but this requires an increase in $a_3$ to $3.6$. The change in $a_3$ also leads to a slight change in $a_1=0.39$ and $\varphi_1=0.73$.

\begin{figure}
\centering
\includegraphics[width=\hsize]{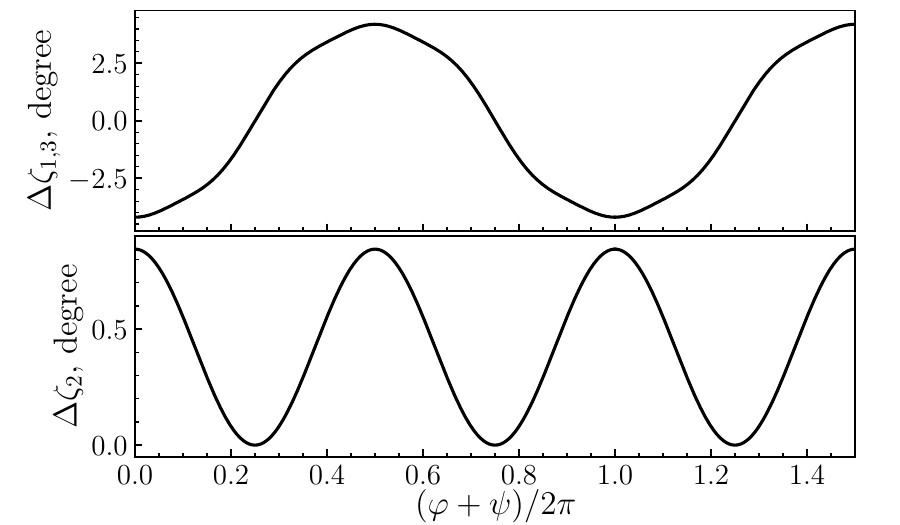}
      \caption{
      The jet deflection angle for the odd powers, $\Delta\zeta_{1,3}$, which manifests itself as nutation, and for the second power, $\Delta\zeta_2$, which manifests itself as variability of the total jet velocity $\beta$. The sign of $\Delta\zeta_{1,3}$ follows the sign of $\mu$.
      }
         \label{fig:ang}
\end{figure}

The observed periodicities do not reflect the change in the angle between the precessional jet direction $\bf j$ and its perturbed direction $\bf j^{\prime}$ itself, but rather the projection of this perturbation onto the line of sight. However, this exact angle has physical meaning and characterizes the net perturbation of the disk by the donor star and gas streams. Expanding the scalar product $\bf j\cdot j^\prime$  up to the quadratic term in $a$ (since the linear terms cancel out), we find that the jet deflection angle is
\begin{equation}
\Delta\zeta_k = a_k\mu^k[1+\sin^2\theta\cos^2(\varphi+\psi-\varphi_k)].
\end{equation}
That is, to within the accuracy of $\sin^2\theta\approx0.1$ the deflection angle is equal to $a(\mu).$
Since $\mu$ varies with the frequency $\omega+\Omega$ ($P=12\fd1$), the angle varies with the same frequency. Figure~\ref{fig:ang} shows the total variation of the angle for the even and odd powers of $\mu$ for the found values of the amplitudes $a_k$.

\section{Conclusion}
\label{sec:conclusion}

In Section~\ref{sect:dev} we confirmed the conclusion of previous works
\citep{1982ApL....23...11K, 2005ApJ...622L.129B} that the jet direction exhibits
random angular fluctuations, independent for each jet, with an amplitude of
$1.\!\!^\circ5$. Against this background, it does not seem surprising that the jet
and counterjet may fail to be parallel. Since the method for determining the full
velocity assumes strict anti-parallelism of the jets, a deviation from
anti-parallelism within such a model biases the full velocity $\beta$.

The random deviations are superimposed on the harmonic variability. We confirm the previously detected periods of $13\fd08$, $11\fd27$, $6\fd29$ and $5\fd83,$ and have also identified a new $3\fd08$ period. The kinematic model of the deviations of the jet and counterjet from the direction set by the precession has allowed us to explain all the detected periodicities. In
this model, the perturbation amplitude of each jet is a function of the cosine
of the angle between this jet and the star--disk radius vector ($\mu_{\pm}$). The
expansion of this function in powers of $\mu_{\pm}$ up to the cubic term yields the
observed set of periodicities in the system, as well as a period of 4.14 days,
which we do not detect, probably due to its low amplitude
($\approx1.3\times10^{-3}$) and the uncertainty in the age of the moving lines,
which hinders the search for short periods.

Comparison of the observed phases of the periodicities with the model predictions
showed that the deviation of the jet $j^{+}$ occurs in the direction of $120^\circ$
from the star--disk radius vector, opposite to the orbital motion, for all the
periodicities except the nutational one ($6\fd29$).

Thus, the analysis of the spectral observations of the moving lines in SS433,
carried out at the Caucasian Mountain Observatory of the Sternberg Astronomical
Institute of Lomonosov Moscow State University in 2020 -- 2026, makes it possible to
explain the periodic deviations from the kinematic model of the source within the
framework of physically justified assumptions about the tidal influence of the
optical component and the dynamical interaction of gas streams with the inclined
precessing accretion disk. The empirically found regularities are important for
understanding the complex physical processes during supercritical accretion onto
black holes in Galactic microquasars and onto supermassive black holes in the nuclei of galaxies.

\begin{acknowledgements}
This work was carried out as part of the State Assignment of Lomonosov Moscow State University. Observations at the instruments and telescopes of the Caucasian Mountain Observatory are supported by the Development Programme of Lomonosov Moscow State University.
\end{acknowledgements}

\section*{Data availability}
Measurements of $z^{\pm}$ are available online.

\begin{contribution}
All authors contributed equally.


\end{contribution}

%
\facilities{CMO:2.5m}

\software{astropy \citep{2013A&A...558A..33A,2018AJ....156..123A,2022ApJ...935..167A},
          }



\bibliography{ss433}{}
\bibliographystyle{aasjournalv7}



\end{document}